\documentclass[a4paper]{llncs}%
\usepackage{multirow}
\usepackage{threeparttable}
\usepackage{color}
\usepackage{colortbl}
\usepackage{geometry}
\usepackage{graphicx}

\definecolor{hellgrau}{rgb}{0.95,0.95,0.95}
\definecolor{dunkelgrau}{rgb}{0.8,0.8,0.8}

\begin{document}

\title{Timing matters: Lessons From The CA Literature On Updating}
\date{April 15, 2010}
\author{Wolfgang Radax\thanks{Corresponding author. Wolfgang Radax' contribution to this paper is funded by research grant \#P19973 of the
Austrian Science Fund FWF. We would like to thank two anonymous referees for their insightful comments.} \and Bernhard Rengs}
\institute{Vienna University of Technology, Austria}
\maketitle

\begin{abstract}
In the present article we emphasize the importance of modeling time in the context of agent-based models. To this end, we present a (selective) survey of the Cellular Automata-literature on updating and draw parallels to the issue of agent activation in agent-based models. By means of two simple models, Schelling's segregation model and Epstein's demographic prisoner's dilemma we investigate the influence of choosing different regimes of agent activation. Our experiments indicate that timing is not a critical issue for very simple models but bears huge influence on model behavior and results as soon as the degree of complexity increases only so slightly. After a brief review of the way commonly used ABM simulation environments handle the issue of timing, we draw some tentative conclusions about the importance of timing and the need for more research towards that direction, similar to the concerted effort on updating in cellular automata.

\end{abstract}

\section{Introduction}
Timing is critical. While this insight is well understood and thoroughly investigated in the cellular automata-literature, we argue that within the ABM community it is treated rather as an orphan. In this article we present a survey of the findings on updating with regard to cellular automata, i.e. when should cells update their states, and then try to translate the lessons learned to the issue of agent activation within agent-based modeling, i.e. in which order should agents be allowed to act.

We would like to make it clear from the outset that though there are activation regimes which seem more fit than others to the modeling of social processes, there is no definite best choice. Rather the choice of the particular activation regime should be informed by the requirements of the process under investigation. This may sound like a common-sense insight. Taking a look at the practical side, i.e. at existing simulation environments, it turns out, however, that certain activation regimes are preset by default or implemented far more easily than others. Given these circumstances, the modeling of timing may often be guided by matters of programming expediency or not even considered explicitly at all.

In the next section we provide a systematic (but far from exhaustive) overview of various updating regimes proposed in the cellular automata literature followed by a presentation of some existing work on activation in agent-based models in Section 3. It turns out that these two issues are highly correlated and insights from the former can be drawn in favor of the latter. We try to illustrate some first tentative conclusions by means of two rather parsimonious, but well-known, models: Schelling's segregation model and Epstein's Demographic Prisoner's Dilemma. In Section 4 we briefly discuss the state of affairs with respect to a number of selected simulation environments (Netlogo, Repast, Repast Simphony) before elaborating on potential implications in the concluding section.

\section{Updating in the CA-literature}
\begin{figure}[ht]
	\centering
		\includegraphics[width=0.70\textwidth]{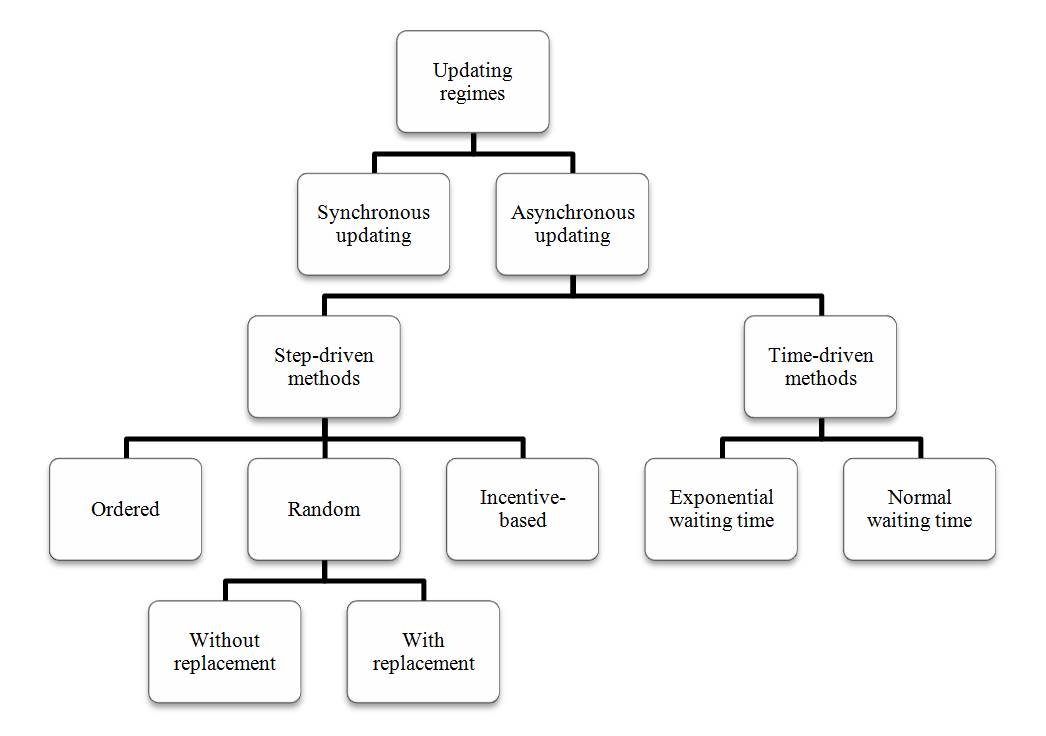}
	\caption{Overview of updating regimes}
	\label{fig:overview}
\end{figure}

For a long time cellular automata were based on the assumption of synchronous updating. Starting in the early 1990s evidence began to amass that many phenomena discovered in cellular automata are a mere artifact of this particular way of handling time and an increasing number of alternatives was to be proposed over the course of the following decade. Figure \ref{fig:overview} provides an overview of the various regimes.

\subsection{Synchronous Updating}
In the beginning there was synchronous updating. Synchronous updating refers to all cells basing their rules at time $t$ on the state of the automaton at time $t-1$. Only after all cells have been active in period $t$, their states are updated. This approach therefore implies a discretization of time and is very reminiscent of difference equation systems. Most of the classics, e.g. Conway's Game of Life or Wolfram's extensive analyses of the properties of cellular automata, make use of this particular updating regime. It is this synchronicity of the individual cells that provided this special regime its name.

This synchronicity also implies that there is some kind of external clock which triggers synchronous events in the process under investigation. While this may be true for some social processes, it is probably rather the exception. Even more important, as Chate and Manneville (cited in \cite{page:incentive}) proclaim, "some of the apparent self-organization in cellular automata is an artifact of the synchronization of clocks." Later work by Huberman and Glance \cite{huberman:async} showed this to be the case in Nowak and May's spatial prisoner's dilemma \cite{nowak:evogames}. As soon as this assumption of a "German marching band model"\footnote{Huberman quoted in \cite{page:computational}} proceeding in lockstep is relaxed, the observed regular patterns of cooperators and defectors break down and the system converges towards total defection in no time.

Taken together this implies that rather than to use synchronous updating by default, the modeler should have good reasons for assuming the system under investigation having an external clock. As an afterthought, if synchronous updating is really the regime of choice then it seems advisable to conduct additional robustness analyses to determine the sensitivity of the model results with respect to this updating regime.

\subsection{Asynchronous Updating}
According to Huberman and Glance \cite{huberman:async}, asynchronous updating can be defined as "choosing an interval of time small enough so that at each step at most one individual entity is chosen [...] to interact with its neighbors. During this update, the state of the rest of the system is held constant. This procedure is then repeated throughout the array for \emph{one player at a time}, in contrast to a synchronous simulation in which all the entities are \emph{updated at once}" (emphases added). Asynchronous updating is therefore more related to continuous time modeling and to a certain degree closer to differential equation systems.

While for synchronous updating the order of activation of the individual cell updates bears no influence on the results, asynchronous updating regimes are highly sensitive to the chosen order. To this end a great number of variations have been proposed, the main distinction being made between step-driven methods and time-driven methods. Given the space limits and the focus of most ABM-simulation environments and indeed many simulation models on a step-driven architecture, we will focus on the former. Suffice it to say here that although time-driven methods, especially utilizing exponential waiting times, seem "the most satisfying updating method from a theoretical point of view" \cite{schoenfisch:updating}, similar behavior can be approximated by using step-driven methods (in this case random asynchronous updating with replacement). For an excellent analysis and comparison of the various methods, the reader is referred to the work of Schoenfisch and de Roos \cite{schoenfisch:updating}.

\subsubsection{Ordered Asynchronous Updating}
The simplest and most straightforward approach to asynchronous updating is to use a predetermined activation order throughout a simulation run, e.g. updating the cells line by line according to their position within the cellular automaton. This approach is known as line-by-line sweep \cite{schoenfisch:updating} or geography-based updating \cite{page:incentive}. Another variant would be to determine a random order at the outset of the simulation run and to use this sequence throughout the whole simulation run. These ordered approaches suffer from a serious drawback, however. It is well acknowledged that they "introduce a lot of additional structure into the automaton" \cite{schoenfisch:updating} by implicitly not treating the cells equally, possibly giving rise to first-mover (dis)advantages.

\subsubsection{Random Asynchronous Updating}
To alleviate the artifacts of ordered asynchronous updating, randomization of the order of activation before \emph{each} period has been proposed. The simpler case is to rearrange the activation order of the cells randomly which amounts to uniform random draws without replacement. Indeed, this approach is the default method used in the most widespread simulation environment Netlogo (when using the ask-primitive) and also encouraged by other simulation environments (Repast, for instance, provides a method for "shuffling" the activation order according to this logic).

An alternative approach is to conduct random draws \emph{with} replacement. This approach requires, however, a definition of the simulation step. Up to here, all the regimes usually assume a simulation step to be finished when all $n$ cells have been updated, i.e. after $n$ updates. The usual way to define a simulation step in the presence of random asynchronous updating with replacement is to consist of $n$ updates, regardless whether all cells have been updated or not. This implies that some cells may have been fired more than once while others where inactive during the simulation step. As has already been mentioned this approach approximates the properties of exponential waiting-times updating, which itself is to be preferred from a theoretical point of view because it imposes the least structure on the output of the automaton.

\subsubsection{Incentive-based Updating}
Page \cite{page:incentive} introduces yet another updating regime which may be of special interest to the social simulation modeler. Instead of assuming a randomly determined updating order, he proposes to activate cells based on their incentives to do so. This regime implies the additional definition of some sort of forward looking utility function in order to derive the utility gain from activation for each cell. The cell which gains the most from firing its rule is the first to do so. Another take along similar lines would be to activate cells according to their discomfort. Imagine, for instance, Schelling's segregation model: instead of activating the agents in a random order, it might be a plausible assumption to let those agents be active first which experience the greatest discomfort in their current situation.

\section{Activation Order in Agent-Based Models}
While the issue of timing has been investigated thoroughly in the cellular automata-literature, the picture looks comparably bleak within the social science ABM-literature. We are aware only of Axtell's work on the effects of interaction and activation \cite{axtell:interaction} and Gilbert's textbook-treatment on the issue of time in ABM \cite{gilbert:abm}. Both works, however, can be mapped neatly to the structure outlined in figure \ref{fig:overview}. For matters of convenience we introduce the term \textbf{activation regime} for the chosen way of activating the agents.

\subsection{Activation Regimes}
Axtell, for instance, compares the effects of using uniform activation and random activation on the results of his agent-based model about the emergence of firms. These activation regimes are related to random asynchronous updating without and with replacement, respectively. Uniform activation refers to each agent being active exactly once per time step, while random activation according to Axtell can be understood as each agent being active exactly once per time step \emph{on average} (i.e. the probability that an agent is activated in the current period is $1/n$, $n$ being the number of agents. Like with random asynchronous updating with replacement, for convenience, it can be assumed that a time step consists of $n$ activations). In the original version \cite{axtell:firms} of his emergence of firms-model, Axtell made use of random activation and achieved a close quantitative fit between the model and empirical data on U.S. firms covering growth rates and firm sizes. When changing this activation regime to uniform activation, the results underwent dramatic qualitative change giving rise to a reversed dependence relation between growth rate and firm size. This serves as an illustrative example of the importance of choosing the adequate activation regime for the modeling problem at hand. Exactly this choice can make the difference between a successful validation and an unsuccessful one.\footnote{Please note that Axtell, of course, didn't choose his activation regime only in order to fit the data but also provided a rationale why exactly this regime is the most suiting for the investigated topic.}

Gilbert distinguishes three broad types of activation order: sequential asynchronous activation, random asynchronous activation, and simulated synchronous activation. Sequential asynchronous activation is related to ordered asynchronous updating as presented in the previous section. Gilbert hastens to add that "[this approach] is rarely a good solution, because the performance of the model may be greatly influenced by the order that is used." In the context of social science simulation we wholeheartedly agree (Cornforth at al. \cite{cornforth:ordered}, however, provide a rationale for using such ordered regimes in the context of modeling, amongst others, computer networks or traffic networks). Random asynchronous activation as described by Gilbert is identical to Axtell's notion of uniform activation. While according to Gilbert this regime is superior to the sequential regime, in his view simulated synchronous activation is the best option at hand. Simulated synchronous activation is related to synchronous updating and implies that "all inputs to agents are completed before all outputs." Based on the insights from the CA-literature we can't agree with this verdict.

\subsection{ABM and Synchronous Activation: An Uneasy Relationship?}
Many agent-based models within social science make use of some sort of asynchronous activation regime. The reason for this bias towards asynchronous regimes is provided by Gilbert as well: implementing synchronous activation is a complicated matter and sometimes it might not even be possible given the requirements of the model. Just think of Schelling's segregation model again (which proves to be a marvelous thinking device in all kinds of circumstances): if all agents would base their decision to relocate in period $t$ on the state of the system in period $t-1$ then you also have to specify some conflict resolution mechanism for the case that two or more agents would like to move to the same spot (having been vacant in period $t-1$). Fates and Chevrier \cite{fates:synchronous} illustrate the intricacies involved in using synchronous updating even with a comparatively simple model.

The problem of model integrity turns up as soon as agents compete against each other for some limited resource with exclusive use. Most often, this will simply be space, when models are based on the assumption that only one agent is allowed per grid cell. The introduction of a conflict resolution mechanism, however, represents an additional source of introducing arbitrariness into the model and increases the degrees of freedom of the model. In many circumstances, there is no real-life precedent for such resolution mechanisms. Such conflict situations often don't arise because many social processes don't proceed in such a synchronous way (of course there are exceptions where an external clock is present or spontaneous self-synchronization takes place). So, the introduction of a synchronous activation regime surely complicates the process of model building and there should at least be a strong reason why the particular phenomenon under investigation is best captured by such an approach. This caution towards synchronicity is emphasized by the findings in the CA-literature that synchronous updating is especially artifact-prone. In our view, synchronous activation regimes are rather the exception from the rule when modeling social processes. As Cornforth et al. \cite{cornforth:ordered} state in the context of cellular automata, "models of multi-agent systems are essentially modeling many processes that occur in parallel, but parallel does not necessarily mean synchronous."

\subsection{Activation Modes}
Another point which merits attention but usually doesn't get it, is the influence of choosing a rule-guided activation mechanism or, alternatively, an agent-guided activation mechanism. We use the term \textbf{activation mode} for this distinction. By agent-guided activation we mean that the first agent fires all rules, then the second agent fires all rules, and so on. Agent-based activation is especially useful when used with asynchronous activation regimes, since this combination is the only way to employ a truly asynchronous activation approach (we will see why in just a minute).

By rule-guided activation on the other hand we understand tat during a simulation step, first all agents fire their first rule, then all agents fire their second rule, then they fire their third rule and so on. This approach seems quite common (see for instance the geopolitical simulations by Cederman \cite{cederman:emergent} \cite{cederman:billard} or Epstein's Demographic Prisoner's Dilemma \cite{epstein:dpd}). It should be noted, however, that this approach introduces a "light version" of the "German Marching Band" through the backdoor by assuming that the modeled process proceeds rule-by-rule, thereby again implying some way of external synchronization. Thus, the modeler implicitly assumes that there is a sub-period in which all agents fire rule 1, then there is a sub-period in which all agents fire rule 2, and so on.

Let's briefly look at the consequences of the possible combinations of activation mode and regime. When using truly synchronous activation, e.g. through buffering the results of the previous time-step (and not using current information about other agents) then both activation modes would lead to the same results, and effectively there would be no sub-periods. Now if the buffering would be updated between the activation of the different rules on the other hand, then sub-periods would form. When we combine rule-based activation with asynchronous activation, we have to settle with a semi-asynchronous form, because the sub-periods are synchronized.

Of course, this notion of sub-periods is very descriptive for certain processes that are for instance season-dependent (farmers bringing in the harvest in fall or animals mating in spring) but there may well be a host of other processes which don't imply such a synchronization.

This is not to argue, of course, pro or against any of these approaches in general. The message is that the modeler should have a good reason why he chooses a particular mode of activation, because this choice may have a big influence on the outcome. Radax and Rengs \cite{radax:repdpd}, \cite{radax:stattest}, for instance, investigated the influence of this choice on the results of the Demographic Prisoner's Dilemma and found them to be highly sensitive to the chosen activation mode. Nevertheless, our impression is rather that such design issues are often times not explicitly based on a rationale with respect to the target process but driven by other reasons, i.e. programming expediency or, in the worst case, mere ignorance.

\section{Exploring the Effects of Activation Regimes}
To illustrate the significance of choosing adequate activation regimes and activation modes, we conducted a number of simulation runs of two well-known agent-based simulation models: Schelling's segregation model and Epstein's Demographic Prisoner's Dilemma. We tried to re-implement these two models faithfully and then we analyzed the effects of choosing uniform and random activation as well as (in the case of the Demographic Prisoner's Dilemma) the effects of activation by rule and activation by agent.

\begin{figure}[ht]
	\centering
		\includegraphics[width=0.40\textwidth]{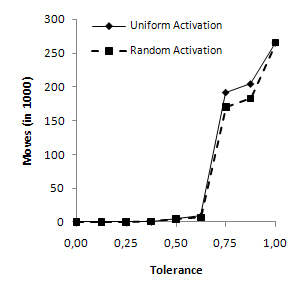}
		\includegraphics[width=0.40\textwidth]{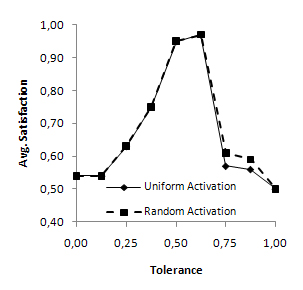}
	\caption{Comparison of the number of moves (left) and average satisfaction (right) in Schelling's segregation model using the movement-rule of Edmonds and Hales with uniform activation and random activation.}
	\label{fig:schelling}
\end{figure}

With respect to the Schelling-model we allowed for two different movement-rules. The first (random everywhere) allows agents to relocate to a random vacant position \emph{anywhere on the grid}. The second movement-rule (Edmonds-Hales) is taken from Edmonds and Hales' reimplementation of the Schelling model in \cite{edmonds:experiment}. This alternative rule lets agents relocate to a random vacant position \emph{within their Moore-neighborhood} only. For both rules we conducted 100 simulation runs with different random seeds for any given value of tolerance (ranging from tolerating zero agents with a different color than the own up to tolerating eight agents with a different color than the own). For each value of tolerance we calculated the average number of moves (each sampled at $t=1000$) as well as the average satisfaction level, i.e. the share of equally-colored agents within the Moore-neighborhood for each agent (again sampled at $t=1000$). Qualitatively, all four simulation experiments (\{random everyhwere, Edmonds-Hales\}x\{uniform activation, random activation\}) reveal the same results. Even closer quantitative scrutiny reveals hardly any differences in the results, especially with the random everywhere-movement rule. In the case of the Edmonds-Hales movement rule, there are some discernible differences in numbers for higher values of the tolerance-parameter (see figure \ref{fig:schelling}).

Of course, these comparatively minor differences are not very supportive of our argument about the importance of activation order, but one has to keep in mind that Schelling's segregation model - for all its brilliance - is as simple as it gets with respect to agent-based models. There is only one rule (movement), the agents are identical (except for their color) and the model is based on weak interaction. According to the definition by Michel et al. \cite{michel:interaction}, "agent actions define a weak interaction when the feasibility of each action's goal does not depend on the action of another agent." To investigate the consequences of the activation regime, we, therefore, turned to a slightly more complex model which is based not on weak, but on strong interaction, defined again by Michel et al. as "actions of agents define a strong interaction when the feasibility of each action's goal depends on the action of another agent." This is surely the case with Epstein's Demographic Prisoner's Dilemma with its reliance on two-person normal-form games.

In the Demographic Prisoner's Dilemma, agents wander around on a torus and interact with other agents by playing a game of Prisoner's Dilemma. Successful agents may give birth to offspring while the unsuccessful ones die and are removed from the simulation. The central rules of the agents are movement, playing, and giving birth to offspring. Because of the space limitations, we refer the reader to \cite{epstein:dpd} for a detailed description of the Demographic Prisoner's Dilemma. Epstein investigated five different settings ordered from a simple baseline setting to increasingly complex settings. In the first setting, agents were not subject to a maximum age. In the second setting, Epstein introduced a maximum age of 100 periods. In settings 3 and 4, Epstein reduced the payoff of mutual cooperation from $5$ (for settings 1 and 2) to $2$ and $1$, respectively. Finally, he reset the payoff of mutual cooperation back to $5$, but introduced a mutation rate of 50 per cent.

\begin{figure}[ht]
	\centering
		\includegraphics[width=0.18\textwidth]{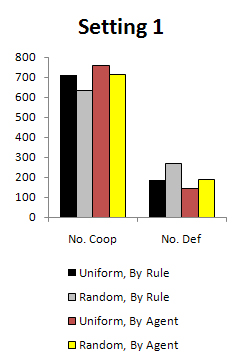}
		\includegraphics[width=0.18\textwidth]{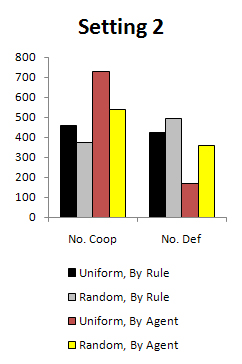}
		\includegraphics[width=0.18\textwidth]{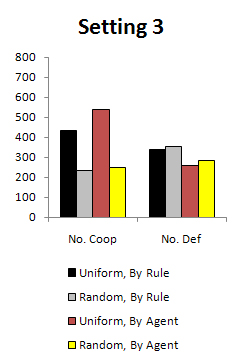}
		\includegraphics[width=0.18\textwidth]{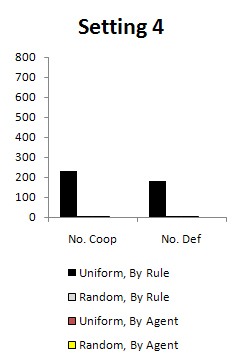}
		\includegraphics[width=0.18\textwidth]{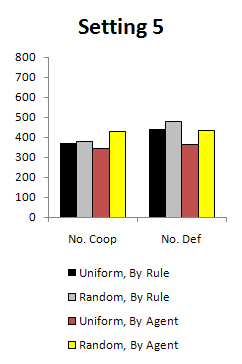}
	\caption{Results of the five settings of Epstein's Demographic Prisoner's Dilemma, dependent on the choice of activation regime and activation mode.}
	\label{fig:epstein}
\end{figure}

We re-ran his five experiments and investigated the influence of choosing between uniform and random activation, on the one hand, and the mode of activation (by agent vs. by rule). Figure \ref{fig:epstein} summarizes our results. Each diagram depicts the number of cooperators and defectors for each of the four experiments (sampled after 1000 periods and averaged over 100 runs with different random seeds) for a given setting. To isolate the impact of the activation regime, comparisons between columns 1 and 2 (uniform vs. random activation when running the simulation "by rule") and between columns 3 and 4 (uniform vs. random activation when running the simulation "by agent") are in order. The influence of the activation mode can be isolated by comparing columns 1 and 3 (activation by rule vs. by agent in the case of uniform activation) and between columns 2 and 4 (activation by rule vs. by agent in the case of random activation).

What appears as striking, is the increasing discrepancy in results for the four experiments along with increasing model complexity. While in the first setting, all four experiments yield similar numbers of cooperators and defectors, the second setting reveals significant quantitative differences when run by "agent". These differences are even magnified in Setting 3 where the outcome is highly sensitive to the chosen activation regime for running the simulation "by rule" as well as "by agent". In Setting 4, also the most volatile in Epstein's original article, the results finally are highly sensitive to the chosen activation regime and activation mode and any change in one of these two brings along qualitative changes in the form of the model population's extinction for almost all runs. Setting 5 breaks this connection between sensitivity and complexity, but we think there's an intuitive reason for this fact. Assuming a mutation rate of 50 per cent introduces a high noise level into the model and probably implies a very low signal-to-noise ratio, or to put it differently: the results in Setting 5 are to a large extent driven only by the random mutation process and not much else.

Although Epstein's Demographic Prisoner's Dilemma exhibits a higher degree of complexity than Schelling's segregation model, it is still a very abstract model of some social process with no directly observable empirical precedent. Looking at models closer to empirically observed phenomena which then have to be validated against data, it soon becomes apparent that - given the impact they produce - the choice of activation regime and activation mode bears a large impact on model results. Therefore, in the process of model verification and validation, the sensitivity of simulations to changes in the regime and mode of activation should be analyzed.

\section{How Are Activation Regimes/Modes Chosen?}
It seems that timing of agent-based simulations is a seldom discussed, but like the last section showed, important topic. Since there is no universally valid method of timing, the question arises how do modelers decide which activation regime to choose. Ideally it should be chosen in way that best represents the described interaction process, though in reality we think that the choice might often be guided by matters of programming expediency. 

The tools of agent-based simulations primarily are the programming language or software package it is planned to be realized with. It is often correctly argued that the results of a simulation do not depend on the language/package used to implement the exactly same model. In reality it shows that it is not easy to come up with the exactly same model using different tools. When ad-hoc programming a simulation using a general-purpose language (using no specific simulation libraries), everything has to be coded anew which requires in-depth knowledge and specific decisions about every simulation aspect and leaves much room for individual mistakes. When using simulation-specific languages or packages, the feature may turn into a problem, since these inherently contain specific assumptions and sometimes even structural decisions about a simulation model. 

Let us illustrate this by taking a very brief look at the current state-of-the-art simulation environments Netlogo, Repast and Repast Simphony. Thinking about the probably most widely available environment Netlogo, one immediately sees that it supports/implicitly encourages the use of one particular activation regime by default. By means of its "ask"-primitive, Netlogo utilizes uniform activation and automatically deals with (unordered) sets of agents, thus avoiding unwanted ordering effects. Now, if one is interested in employing some ordered activation regime or random activation, one has to go some extra miles and make use of lists, thereby loosing much of Netlogo's advantage of simplicity.

Repast, on the other hand, forces the modeler to explicitly contemplate about the activation regime. By default, no randomization of the agent-list is conducted and it is at the modeler's discretion how to randomize the activation order (or to simply forget about it). Admittedly, Repast is biased towards uniform activation as well, for it provides a standard-routine for this approach but not for random activation. Repast Simphony offers a similar array of options, but if one wants to use other activation regimes than ordered ones and uniform activation, additional programming is necessary.

While especially Netlogo cannot be applauded enough for making accessible the fascination of agent-based modeling to a large audience, at the same time this easy access approach, in our view, bears part of the blame for the negligence of the potential implications of modeling time.

\section{Conclusion}
The central point we tried to convey in this article is the importance of timing in agent-based simulation models and the simultaneous negligence of this importance. This situation can well be compared with the state of the Cellular Automata literature prior to the extensive studies about the role of updating. Prior to these studies, synchronous updating was simply assumed as a convention and not further questioned. Only through the concerted efforts of the update studies cited throughout this text was the importance and the sometimes dramatic influence of timing understood better. We think that a similar discussion is warranted for agent-based modeling as well and tried to sketch a few threads along which such a discussion could proceed.

On the one hand, the CA literature has shown a rich array of possible updating regimes and their theoretical properties. These insights could without much effort be translated to ABM as well and show alternatives to the usual mode of conduct (which is most often uniform activation). On the other hand, there is also the issue whether to conduct agent-guided or rule-guided activation which bears great influence on model results. A third potential and closely related thread which we couldn't touch because of space limitations is the assumption of most agent-based models of a fixed order of rules (irrespective of the chosen activation mode). As a matter of convenience (and, of course, clarity, we hasten to add), simulation models assume some "innate" order of rules: First Rule A, then Rule B, then Rule C, and so on. But there are alternatives to this approach as Ruxton and Saravia \cite{ruxton:order} demonstrate for cellular automata. They introduce alternative orders of rules. In the context of, for instance, Epstein's Demographic Prisoner's Dilemma, it might as well be plausible instead of assuming the fixed order of movement-playing-giving birth to assume any other order of rule activation or to simply randomize it.

Of course, there is a rationale behind choosing a simple and straightforward mechanism for activation. Keeping things simple allows the modeler to focus on the hypothesized key relations of the target process. But if the choice of activation mechanism drives large parts of the results then a more detailed discussion of this issue is certainly in order. We believe such a discussion of the timing issue to be a potentially very fruitful one. And although we can't estimate what such discussions could unveil, we are quite sure, drawing on the CA-updating-literature, of the central conclusion to be gained after intensive investigation: There is no one-size-fits-all solution to modeling time, but rather there is only a best way of doing it for each particular social process to be modeled. Rather than choosing the most convenient way of realizing this issue or following some convention of how it is usually done, it is the modeler's duty to choose the one mode of activation that fits the problem at hand the best.

\bibliographystyle{splncs}

\end{document}